\documentclass{aa}
\usepackage{psfig}
\usepackage{epsfig}

\def \hcm {\hbox {\ifmmode $ cm$^{-2}\else cm$^{-2}$\fi}}
\def \arcmin {\hbox{$^\prime$}}
\def \arcsec {\hbox{$^{\prime\prime}$}}

\def\approxgt{\mathrel{\hbox{\rlap{\lower.55ex \hbox {$\sim$}}
        \kern-.3em \raise.4ex \hbox{$>$}}}}
\def\approxlt{\mathrel{\hbox{\rlap{\lower.55ex \hbox {$\sim$}}
        \kern-.3em \raise.4ex \hbox{$<$}}}}

\begin{document}

\title{XMM-Newton observation of the brightest X-ray flare detected so far from Sgr\,A*  }

\author{D. Porquet\inst{1} 
        \and P. Predehl\inst{1}  
         \and B. Aschenbach\inst{1}
           \and N. Grosso\inst{2}
              \and A. Goldwurm\inst{3}
                \and P. Goldoni\inst{3}
                 \and R.S. Warwick\inst{4}
                \and A. Decourchelle\inst{3}
       }
\offprints{Delphine Porquet\\ (dporquet@mpe.mpg.de)}

\institute{Max-Planck-Institut f\"{u}r extraterrestrische Physik,
P.O. Box 1312, Garching bei M\"{u}nchen D-85741, Germany
\and Laboratoire d'Astrophysique de Grenoble, Universit\'e Joseph-Fourier, BP53, 38041 Grenoble Cedex 9, France
\and CEA, DSM, DAPNIA, Service d'Astrophysique, C.E. Saclay, 91191 Gif-Sur-Yvette Cedex, France
\and Department of Physics and Astronomy, University of Leicester, Leicester LE1 7RH, UK 
}
\date{Received ...; Accepted ...  }

\abstract{We report the high S/N observation on October 3, 2002 with {\sl XMM-Newton} 
of the brightest X-ray flare detected so far from \object{Sgr\,A*} 
 with a duration shorter than one hour  ($\sim$ 2.7\,ks).
 The light curve is almost symmetrical with respect to the peak flare,   
 and no significant difference between the soft and hard X-ray range  is detected. 
The overall flare spectrum is well represented by an absorbed power-law 
 with a soft photon spectral index of $\Gamma$=2.5$\pm$0.3, 
 and a peak  2--10\,keV luminosity of 
 3.6$^{+0.3}_{-0.4}\times$10$^{35}$\,erg\,s$^{-1}$, 
i.e. a factor 160 higher than the Sgr\,A* quiescent value.   
 No significant spectral change during the flare is observed.
  This X-ray flare is very different from other bright flares 
 reported so far: it is much brighter and softer.   
 The present accurate determination of the flare characteristics  
  challenge the current interpretation of 
 the physical processes occuring inside the very 
 close environment of Sgr\,A* by  
 bringing very strong constraints 
 for the theoretical flare models.
\keywords{Galaxy: center -- X-rays: individuals: Sgr\,A* -- X-rays: general -- Radiation mechanisms: general }
}
\titlerunning{The brightest X-ray flare detected so far from Sgr\,A*}
\authorrunning{Porquet et al.}
\maketitle

\section{Introduction}
 Measurements of star motions in the central part of our Galaxy 
  demonstrated the presence of 
a high concentration of matter within a very compact
region of 0.01 pc at a distance of 8 kpc. These results prove the existence
of a central supermassive black hole of a mass of about 3$\times$10$^{6}$\, M$_{\odot}$ 
at the dynamical center of our Galaxy (e.g., Sch{\" o}del et al. \cite{Schodel2002}), 
 coincident with the compact radio source Sgr A*.
  Surprisingly this source is much fainter than expected 
from accretion onto  a super-massive black hole.   
 In particular  in the 2--10\,keV energy band its X-ray luminosity 
 is  only about  2.2$\times$10$^{33}$\,erg\,s$^{-1}$ 
 within a radius of 1.5$^{\prime\prime}$  
 (Baganoff et al. \cite{Baganoff2003}, hereafter B03).
 This value may in fact be considered as an upper limit since 
 this region contains other components such as stars 
 (e.g., Eckart \& Genzel \cite{EG97}), hot gas, etc.   
 Thus, Sgr\,A* radiates in X-rays at about 11 orders 
of magnitude less than its corresponding Eddington luminosity. 
 Its bolometric luminosity is only
   about 3$\times$10$^{-8}$ L$_{\rm Edd}$ (Melia \& Falcke \cite{MF2001}, 
Zhao et  al. \cite{Zhao2003}).
 This has motivated the development of various radiatively inefficient
accretion models to explain the dimness of the Galactic Center black hole, 
e.g. Advection-Dominated Accretion Flows 
(e.g.,  Narayan et al. \cite{Narayan98}), 
jet-disk models (e.g., Falcke \& Markoff \cite{Falcke2000}),
 Bondi-Hoyle with inner Keplerian flows (e.g., Melia et al. \cite{Melia2000}). 
 The recent discovery of X-ray flares from Sgr A* has provided new 
exciting perspectives for the understanding of the processes at work in the 
galactic nucleus. 
The first detection of such events was found with {\sl Chandra} in October 2000.
The flare has a duration of  
 about 10\,ks, with  L(2--10\,keV)= 1.0$\pm$0.1 $\times$ 10$^{35}$\,erg\,s$^{-1}$ 
 for the flare peak, i.e. about 45 times the quiescent state.
 (Baganoff et al. \cite{Baganoff2001}, hereafter B01).   
  The photon flare spectral index, obtained from the simultaneous 
fitting of the quiescent phases and the flare phase, was $\Gamma$=1.3$^{+0.5}_{-0.6}$
 (at 90$\%$ confidence level). 
 A second significant  flare was detected with {\sl XMM-Newton} 
 by Goldwurm et al. (\cite{Goldwurm2003}) who found a monotonic 
 flux rise up to a factor of about 20--30 in the last 900\,s of the observation. 
 They found also a rather hard spectrum with $\Gamma$=0.9$\pm$0.5 
 (at 68.5$\%$ confidence level), 
 with a maximum observed 2-10\,keV luminosity 
 of about 6$\times$10$^{34}$\,erg\,s$^{-1}$. 
 Baganoff et al. (\cite{Baganoff2003b}) estimated, with the analysis of 
several {\sl Chandra} observations, that  
 X-ray flares happen at a rate of 1.2$\pm$0.6 per day.\\
\indent We report here the detection and the analysis 
 of the up-to-now brightest  X-ray flare  from SgrA*, observed 
 in October 2002 with {\sl XMM-Newton}. 
The present data provide the highest signal to noise measurements 
 of a Sgr A* X-ray flare, reported so far.
\section{XMM-Newton observation and results}
 
$~$ The Galactic Center was observed on October 3, 2002 
with Sgr\,A*   at the center of the {\sl XMM-Newton}/EPIC field of view  
 (see Turner et al. \cite{T2001} and Str{\" u}der et al. \cite{S2001}  
 for the description of the EPIC cameras MOS and PN, respectively). 
The MOS  (exposure time $\sim$16.8\,ks) and PN ($\sim$13.9\,ks) 
 cameras were both operated 
in the prime full window mode, with the medium 
and the thick filter selected, respectively. 
The data were processed with the 
{\sc XMM Science Analysis Software} (version 5.4.1). 
 X-ray events with pattern 0--12 and 0--4 are used 
 for the MOS and PN, respectively. \\
\indent We find a strong increase of the count rate during about 3\,ks. 
To identify the counterpart of this X-ray flare, 
we improve the astrometry by cross-correlating the 8 PN sources detected 
in the 0.5--1.5\,keV energy band during the whole observation 
with the 2MASS All-Sky Point Source Catalog using the {\tt eposcorr} SAS task. 
We find a reliable 2MASS counterpart for each of these 
serendipitous PN sources, from which three are Tycho sources, 
by introducing boresight correction and rotational offset of 
(-1.2\arcsec,-0.5\arcsec), and 2.7\arcmin, respectively. 
The refined position of the X-ray flaring source is then  
$\alpha_{\rm J2000}$= 17$^{\rm h}$45$^{\rm m}$39.98$^{\rm s}$,
$\delta_{\rm J2000}$= $-$29$^{\circ}$00$^{\prime}$28.4$^{\prime\prime}$, 
with a 90\% confidence level error box of 2.3\arcsec.
 The radio counterpart of Sgr\,A* at  
$\alpha_{\rm J2000}$=$17^{\rm h}45^{\rm m}40.0383^{\rm s}\pm0.0007^{\rm s}$,
$\delta_{\rm J2000}$=$-29^{\circ}00^{\prime}28.069^{\prime\prime}\pm0.014^{\prime\prime}$ 
(Yusef-Zadeh et al. \cite{Yusef99}), is located 
only 0.8\arcsec~ away 
from the flaring source. Since no other X-ray source was found by {\sl Chandra}  
in our error box, we conclude that the X-ray flare is very likely to be associated with Sgr\,A*. 
 For the light curves and spectral analysis,  we use an extraction region 
 of 10$^{\prime\prime}$ radius around Sgr\,A*, where the solar flare
 background  is negligible with a maximum count rate of 
 6$\times$10$^{-3}$\,cts\,s$^{-1}$. 

\begin{figure}[h!]
\includegraphics[height=9.5cm,width=8cm]{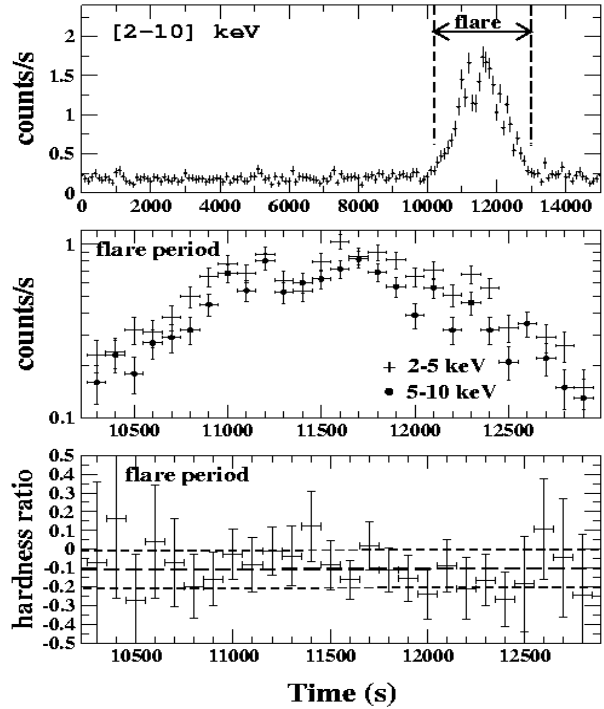}\\
\vspace*{-0.56cm}
\caption{EPIC light curves (MOS\,1+MOS\,2+PN)  and hardness ratio (HR) 
 within a radius of 10$^{\prime\prime}$ around Sgr A* position. 
 The  time binning is 100s, and the error bars indicate 1$\sigma$ uncertainties.
 {\it Upper panel}: The 2--10\,keV light curve  
 shows the quiescent and the flare periods.  
{\it Middle panel}: The 2--5\,keV and 5--10\,keV  light curves  during  the flare period.
 {\it Lower panel}: quiescent subtracted HR during the flare period.  
 Long dashed line: mean HR value, 
 and short dashed lines: 1$\sigma$ error bars. 
\vspace*{-0.4cm}
}
\label{fig1}
\end{figure}

\subsection{The EPIC light curve}

The EPIC light curve in the 2--10\,keV energy range shows 
 for the first 10\,ks a quiescent period and 
then a very sharp and intense 
flare with a duration shorter than one hour 
($\sim$2.7\,ks, see Fig.~\ref{fig1} {\it top panel}). 
The flare shape is almost symmetric relative to the peak. 
 Both the rise and decay phases can be fitted by 
 an exponential with characteristic times  of 610\,s and 770\,s, respectively. 
We also observe  a ``micro-structure'' at about 11,200\,s 
 with a significance of at least 3$\sigma$ 
 on a time-scale as short as 200\,s, which
  can further constrain the emitting region to a size of 
about 7 Schwarzschild radii for the black hole mass 
 of about 3$\times$10$^{6}$\,M$_{\odot}$. 
  The 2--10\,keV count rate 
 at the flare peak  is 1.74$\pm$0.13 cts\,s$^{-1}$,
  i.e. 9.2$^{+3.3}_{-2.2}$ times higher than the mean value 
 measured during the quiescent period  (0.19$\pm$0.04 cts\,s$^{-1}$).
 B03 using {\sl Chandra} found that the inner 10$^{\prime\prime}$ radius region 
around Sgr\,A* contains six other X-ray point sources as well as  a  diffuse component, 
 and estimated that the flux from Sgr\,A* during its quiescent state   represents about 10$\%$ 
 of the total 2--10\,keV flux. 
Therefore we infer for the present flare a count rate increase of a factor of about  90  
between the quiescent state of Sgr\,A* and the flare peak. 
The shapes of the light curves in the {\sl soft} (2--5\,keV) 
 and {\sl hard} (5--10\,keV) 
 energy bands are very similar (Fig.~\ref{fig1}, {\it middle panel}). 
  We report in  Fig.~\ref{fig1} ({\it lower panel}), 
 the hardness ratio (HR)  during the flare period: 
 HR=(hard$-$soft)/(hard$+$soft). 
 We removed the quiescent level contribution, 
assuming that it is constant during the flare:  
mean count rates of 0.12$\pm$0.03 cts\,s$^{-1}$, 
 and  of 0.07$\pm$0.03 cts\,s$^{-1}$, respectively for  
the soft and hard  energy bands. 
 This corresponds to a HR value for the quiescent period 
of $-$0.29$\pm$0.02. 
 We find a  flare HR of  $-$0.1$\pm$0.1,   
  indicating a rather soft spectrum.  
  No significant spectral evolution is observed
 between the [2--5]\,keV [5--10]\,keV ranges 
 (a similar result is found for a different energy range, e.g. [2--3.5]\,keV).
 We find an indication that the flare presents a higher  
 HR value, compared to the quiescent, 
in a restricted  time period of the rise phase  
  (i.e 11,000 $\leq$ time $\leq$ 11,400\,s). 
 However, this does not imply a change
in spectrum of the flaring source since  
the emission within 10$^{\prime\prime}$ 
during the quiescent period is not dominated by Sgr\,A*.

\subsection{Spectral analysis of the X-ray flare}

\begin{figure}[h!]
\includegraphics[angle=-90,width=\columnwidth]{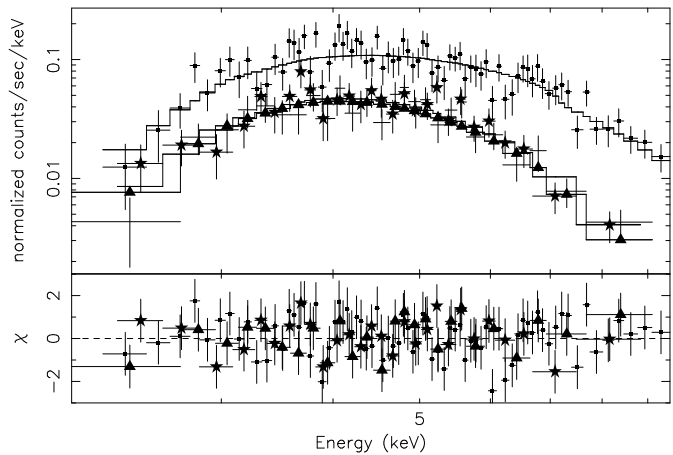}
\caption{EPIC spectra of the flare, within a radius of 
10$^{\prime\prime}$ around Sgr A* position.  
 The spectrum for the quiescent period is used as background. 
A power-law model has been fitted taking into account ISM 
absorption and dust scattering. See Table~\ref{table:fit} for parameter fit values.}
\label{fig:fitpl}
\end{figure}

Fig.~\ref{fig:fitpl} shows the extracted spectra for the overall flare period. 
 The overall flare spectra and the rise and decay phase spectra 
 were binned to a minimum of  
20 counts and 10 count per bin, respectively. 
{\sc xspec v11.2.0} is used for the spectral analysis.
 We use the response matrices from the {\sc rmfgen} and {\sc arfgen} 
SAS tasks. The errors and upper limits 
quoted correspond to 90$\%$ confidence level  for one interesting parameter. 
 The quiescent spectrum  including the contribution of Sgr\,A* in 
 its quiescent phase and the contribution of the other sources 
  in a radius of 10$^{\prime\prime}$ around Sgr\,A*  
  is subtracted from the flare spectrum.  
  We fit the data with an absorbed power-law model taking into account 
the dust scattering effects along the line-of-sight, 
 using the {\sc scatter} model soon available in {\sc xspec} 
 (Predehl \& Schmitt \cite{PS95}).  
  We fix for the {\sc scatter} model,
   a visual extinction value A$_{\rm V}$ of 30\,mag, 
 as determined from IR observations of stars 
close to Sgr\,A* (e.g., Rieke et al. \cite{Rieke89}). 
 The interstellar medium (ISM) column density 
 is fitted using the model {\sc tbabs/xspec} (Wilms et al. \cite{Wilms2000}), 
 which includes updated cross-sections for X-ray absorption by the ISM.  
 We also use their revised ISM abundances, which produce 
 changes in ISM cross-sections of up to 30$\%$ with respect to
 solar abundances  generally used as reference abundances for the ISM.  
 This means that the absorption column values 
 found here are about 30$\%$ higher than the ones  
  obtained assuming solar abundances. 
\begin{table}
\caption{Fit of the EPIC flare spectra, in the 2--10\,keV energy range, 
with absorbed power-law (pow),  bremsstrahlung (brems), 
 black-body (bb) or mekal models, taking into account dust scattering. 
 Frozen parameters are given between brackets.
(a): ${\cal N}_{H}$ is units of 10$^{23}$\,cm$^{-2}$.
 (b): Z is the metal abundance relative to the ISM one.  
(c): The fluxes are unabsorbed and expressed in units 
 of 10$^{-11}$erg\,cm$^{-2}$\,s$^{-1}$.
}
\begin{tabular}{ccc@{\ }c@{\ }cc@{\ }c@{\ }}
\hline
\hline
\noalign {\smallskip}
Model   & ${\cal N}^{\rm (a)}_{H}$  & $\Gamma$/kT  & Z$^{\rm (b)}$        &F$^{\rm (c)}_{2-10 {\rm keV}}$   & $\chi^{2}$/d.o.f. \\
\noalign {\smallskip}
\hline
\noalign {\smallskip}
pow          &   2.0$\pm$0.3     &  2.5$\pm$0.3  & [1.0]   & 2.4$^{+2.0}_{-1.0}$ &     102.4/123  \\
\noalign {\smallskip}
pow          &  1.1$\pm$0.1     &   [1.3]     & [1.0]   &  1.6$\pm$0.2           &    149.0/124 \\
\noalign {\smallskip}
brems       &   1.7$\pm$0.2      &  6.1$^{+2.3}_{-1.3}$ &  [1.0] & 2.0$^{+0.6}_{-0.4}$ & 104.3/123     \\
\noalign {\smallskip}
bb       &   1.0$\pm$0.2      &  1.47$^{+0.12}_{-0.06}$ &  [1.0] & 1.3$\pm$0.1 & 113.6/123     \\
\hline
\noalign {\smallskip}
mekal       &   1.7$\pm$0.2       &  5.7$^{+2.3}_{-1.2}$ & $<$0.17 & 2.0$^{+0.6}_{-0.5}$ &    103.8/122     \\
\noalign {\smallskip}
mekal       &   1.1$^{+0.2}_{-0.1}$ &  27$^{+30}_{-13}$ &  [1.0]  & 1.6$\pm$0.2         &    147.9/123     \\
\noalign {\smallskip}
mekal       &   1.0$\pm$0.1       &  $\geq$48                    &    [2.0]       & 1.6$\pm$0.2         &    159.7/123   \\
\noalign {\smallskip}
\hline
\hline
\end{tabular}
\label{table:fit}
\end{table}
 The power-law continuum model  with a soft photon index 
 gives a good representation of the data (see Fig.~\ref{fig:fitpl} and Table~\ref{table:fit}),
 and we find at the flare peak a 2--10\,keV luminosity 
of 3.6$^{+0.3}_{-0.4}\times$10$^{35}$\,erg\,s$^{-1}$, assuming d=8\,kpc. 
 Fixing  $\Gamma$=1.3, as found in B01, we obtain 
 a non satisfactory data fit (Table~\ref{table:fit}). 
 The presence of  narrow Fe\,K$_{\alpha}$ Gaussian emission lines 
($\sigma$=10\,eV)  are not statistically required   
 ($\Delta \chi^{2}<$1 for one additional parameter). 
  We find at 6.4\,keV, 6.6\,keV,  6.7\,keV,  and 7.0\,keV 
  upper limits on the equivalent widths of   
 33\,eV, 122\,eV, 123\,eV, and 65\,eV, respectively. 
 Likewise the presence of a broad emission line near 6.7\,keV 
 ($\sigma$=250\,eV) and an absorption edge at 7.5\,keV 
 are  not statistically required and we obtain EW$<$173\,eV and $\tau<$0.2, 
 respectively.  
  The data are also well fitted by a bremsstrahlung continuum model 
with a temperature of about 6.1\,keV (Table~\ref{table:fit}). 
 We also test an  optically thin plasma  using the {\sc mekal} model 
with ISM abundances, letting the metal abundance vary as a free parameter, 
we find a good fit  for  a  relative metal underabundance of less than 0.17. 
 While fixing the relative metal abundance to be equal or two times higher 
 than the ISM abundances,  the data fit is not satisfactory, with very high temperatures 
 and therefore a large over-prediction of a H-like iron line at 7\,keV (Table~\ref{table:fit}). 
Surprisingly, the spectrum can also be well fitted by a black-body with 
kT=1.47\,keV with no significant change of temperature  all over the flare. 
 The apparent size of the emitting area is about 10\,km$^{2}$ 
at the peak  for a distance of 8\,kpc. 
  The spectral fits of the rise and decay phases of this flare 
 are statistically consistent with each other with no change of the photon index 
 (${\cal N}_{H}$ fixed to 2$\times$10$^{23}$\,cm$^{-2}$),
  with $\Gamma$=2.4$\pm$0.2 ($\chi^{2}$/d.o.f.=103.0/122), 
and $\Gamma$=2.5$\pm$0.2 ($\chi^{2}$/d.o.f.=131.9/126), 
respectively.  This confirms the lack of significant hardening 
according to the HR found. 

\section{Discussion}

\indent Several models have been proposed to explain the 
 first Sgr\,A*  X-ray flare observed with {\sl Chandra}. 
 For instance, it could be produced by a sudden 
 enhancement of accretion through the inner hot 
 circularized flow  predicted by the model of Liu \& Melia (\cite{Liu2002}).  
 When the viscosity increases, synchrotron self-Compton (SSC) 
 of the millimeter to sub-millimeter photons is the dominant process, 
 and a soft spectrum is expected.   
  Such flares would be coupled with a similar increase in the mm 
 and sub-mm ranges, which have not been yet observed 
 (e.g., from March 2001 to May 2002, Zhao et al. \cite{Zhao2003}). 
  If the viscosity decreases then thermal Bremsstrahlung is
the dominant process, 
and a hard spectrum would be generated, consistent with the slopes 
observed in the two other X-ray (much lower S/N)  flares. 
   Three possible flare mechanisms have been proposed by Markoff et al.
 (\cite{M2001}) on the basis of a jet model: 
 increased jet power or accretion rate, increased heating 
of  quasi-thermal relativistic particles, or sudden shock acceleration. 
 The first scenario predicts a steep spectral index, 
  and a huge radio flare for an accretion rate increase, 
 and simultaneous flaring at all frequencies for an jet power increase, 
 but this have never been yet observed.  
 The second and third scenarii predict harder spectrum and 
 were compatible with the spectral index of the first {\sl Chandra} X-ray flare.\\
\indent  The flare luminosity excludes normal stars 
 as the origin of this event, but compact objects,
 although unlikely, cannot be totally excluded
(see discussion in B01).
  We rule out the possibility that the symmetrical shape of 
the flare light curve  
 is produced by  micro-lensing of a star located behind Sgr\,A*.
 Indeed from Zinnecker (\cite{Zinnecker2003})'s formulae, 
  we find that hour duration micro-lensing events imply  
 very high tangential velocities for the lensed star 
 and unphysical distances from the black hole, less than 
 one Schwarzschild radius. For instance a velocity as high as 
 10,000\,km\,s$^{-1}$ corresponds to a distance of about only 10$^{12}$\,cm. 
  The duration of the trailing edge of this flare is  very short. 
  If the physical process is thermal, 
 this could be explained by the very rapid cooling of a  plasma 
  not magnetically confined, as proposed by  Merloni \& Fabian (\cite{Merloni2002}).\\
\indent  The total column density inferred 
from the spectral fit is about 1.7--2.0$\times$10$^{23}$\,cm$^{-2}$
 for the power-law, bremsstrahlung and mekal models.
 The visual extinction A$_{\rm V}\sim$30\,mag, inferred from IR observations of 
 stars very close to Sgr\,A* (Rieke et al. \cite{Rieke89}),
 corresponds to an absorption column density 
 along the line-of-sight of about 6$\times$10$^{22}$\,cm$^{-2}$ (Predehl \& Schmitt \cite{PS95}). 
 This discrepancy between A$_{\rm V}$ and  ${\cal N}_{H}$ 
 (also reported by Predehl \& Tr\"{u}mper \cite{Predehl94})  
  may indicate an additional, probably related to the source, 
gaseous absorption  from nearly neutral gas.  
  A spectral analysis of other X-ray sources 
in the {\sl XMM-Newton} field of view  
 will enable to check whether this absorption excess 
 is peculiar to Sgr\,A* (Predehl et al., in preparation). 
 Nayakshin et al. (\cite{NCS2003}) proposed 
  that such X-ray flares are due to the passage of stars through the accretion 
 disk around Sgr\,A*, and that  extra absorption may be related to 
 the neutral accretion disk photosphere 
 through which the star passes.  
The excess absorption could also be explained by an overabundance of heavy 
 metal elements  from a putative Supernova remnant 
 or  He and other heavy elements from 
 He stars located in the neighbourhood of Sgr\,A* (e.g., Eckart et al. \cite{E95}). 
 However, for the blackbody model no extra absorption is required. \\
\indent  We find also a soft X-ray flare spectrum 
 with $\Gamma$=2.5$\pm$0.3,   
much softer than for the two other bright flares reported up to now 
 (with lower S/N): 
 $\Gamma$=1.3$^{+0.5}_{-0.6}$ (B01), and  
$\Gamma$=0.9$\pm$0.5 (Goldwurm et al. \cite{Goldwurm2003}).   
 The present spectral index is  consistent with the ones
  found during the quiescent periods of Sgr\,A* by {\sl Chandra}: 
$\Gamma$=2.5$^{+0.8}_{-0.7}$ in September 1999 (B03), 
 and $\Gamma$=1.8$^{+0.7}_{-0.9}$ in October 2000 (B01). 
 Such a soft X-ray spectrum is predicted by the scenario 
 proposed by  Liu \& Melia (\cite{Liu2002}) in case of an increased viscosity 
 of the accretion flow,  and by the one proposed by  Markoff et al.
 (\cite{M2001}) in case of an increase of accretion rate or jet power.
 However the predictions of these two latter models are not
 consistent with observational constraints found at other wavelengths 
  (e.g, Zhao et al. \cite{Zhao2003}, Hornstein et al. \cite{Hornstein2002}). 
 Although, the Nayakshin et al. (\cite{NCS2003})  star-disk interaction model 
 predicts a soft X-ray spectral index ($\Gamma\sim$2.3) and a moderate 
 Fe K$_{\alpha}$ line EW of about 60\,eV
  the allowed temperature values (kT=1 and 70\,keV) are very fine-tuned  
 (S. Nayakshin, private communication).
 Emission from a  hot optically thin plasma with kT$\sim$5.7\,keV 
 and a low relative metal abundance ($<$0.17 compared with the ISM abundances)
 is also consistent with the observed spectrum. 
 But the high metal depletion is difficult to explain 
 in view of the ambient, diffuse and metal rich emission and the presence  
 of close supernova remnants such as Sgr\,A East.  
 If the flare is related to the accretion disk, the low metallicity could indicate a 
 less evolved material in the disk. \\
\indent In conclusion, the present observation
  strongly constrains the flare modeling of Sgr\,A*.
 
\begin{acknowledgements}
This work is based on observations obtained with {\sl XMM-Newton}, 
an ESA science mission with instruments and contributions directly 
funded by ESA Member States and the USA (NASA).  
 We would like to thank the referee, F.K. Baganoff, for constructive 
 comments and suggestions. 
D.P. is supported by a MPE fellowship.
\end{acknowledgements}



\begin{thebibliography}{}

\bibitem[2001]{Baganoff2001} 
Baganoff, F.~K., Bautz M.~W., Brandt W.~N.,  et al. 2001, Nature,  413, 45 (B01)
\bibitem[2003a]{Baganoff2003}
Baganoff, F.~K., Maeda Y., Morris M., et al. 2003a, ApJ, 591, 891 (B03)
\bibitem[2003b]{Baganoff2003b}
Baganoff, F.~K, Bautz, M. W., Ricker, G. R. et al. 2003b, Astron. Nachr., Vol. 324, in press 

\bibitem[1995]{E95} 
Eckart, A., Genzel, R., Hofmann, R., et al. 1995, ApJ, 445, L23 

\bibitem[1997]{EG97} 
Eckart, A. \& Genzel, R.\ 1997, MNRAS, 284, 576 

\bibitem[2000]{Falcke2000} 
Falcke, H. \& Markoff, S. 2000, A\&A,  362, 113 

\bibitem[2003]{Goldwurm2003} 
Goldwurm, A., Brion E., Goldoni P., et al. 
2003, ApJ, 584, 751

\bibitem[2002]{Hornstein2002} 
Hornstein, S.~D., Ghez A.~M., Tanner A., 
et al. 2002, ApJ,  577, L9 

\bibitem[2002]{Liu2002} 
Liu, S. \& Melia, F. 2002, ApJ,  566, L77 

\bibitem[2001]{M2001} 
Markoff, S., Falcke H., Yuan F., 
 et al. 2001, A\&A,  379, L13 

\bibitem[2001]{MF2001} 
Melia, F. \& Falcke, H.\ 2001, ARAA, 39, 309 
\bibitem[2000]{Melia2000} 
Melia, F., Liu, S. \& Coker, R. 2000, ApJ,  545, L117 

\bibitem[2002]{Merloni2002} 
Merloni, A. \& Fabian, A.C. 2002, MNRAS,  332, 165 

\bibitem[1998]{Narayan98} 
Narayan, R., Mahadevan R., Grindlay J.~E., 
 et al. 1998, ApJ,  492, 554 

\bibitem[2003]{NCS2003}
Nayakshin, S., Cuadra J., \& Sunyaev R. 2003, A\&A, submitted [astro-ph/0304126]

\bibitem[1994]{Predehl94} 
Predehl, P. \& Tr\"{u}mper, J. 1994, A\&A,  290, L29 
\bibitem[1995]{PS95} 
Predehl, P. \& Schmitt, J.H.M.M. 1995, A\&A,  293, 889 

\bibitem[1989]{Rieke89} 
Rieke, G.H., Rieke, M.J. \& Paul, A.E.\ 1989, ApJ, 336, 752 

\bibitem[2002]{Schodel2002} 
Sch{\" o}del, R., Ott, T., Genzel, R., et al.\ 2002, Nature, 419, 694 

\bibitem[2001]{S2001} 
Str{\" u}der, L., Briel, U., Dennerl, K., et al.\ 2001, A\&A, 365, L18 

\bibitem[2001]{T2001} 
Turner, M.~J.~L., Abbey, A., Arnaud, M., et al.\ 2001, A\&A, 365, L27 

\bibitem[2000]{Wilms2000} 
Wilms, J., Allen, A. \& McCray, R. 2000, ApJ,  542, 914 

\bibitem[1999]{Yusef99} 
Yusef-Zadeh, F., Choate, D., \& Cotton, W.\  1999, ApJ, 518, L33 

\bibitem[2003]{Zhao2003} 
Zhao, J. et al., Young, K.~H., Herrnstein, R.~M., et al. 
2003, ApJL, 586, L29 

\bibitem[2003]{Zinnecker2003}
Zinnecker, H. 2003, IAU, vol. 21, in press [astro-ph/0301074]


\end{thebibliography}
           \end{document}